\documentclass[prb,twocolumn,
superscriptaddress,
noeprint, 
nourl,
showpacs,
aps
]{revtex4-2}
\usepackage{graphicx}
\usepackage{amsmath}
\usepackage{amsfonts}
\usepackage{amssymb}
\usepackage[colorlinks=true,linkcolor=blue,urlcolor=blue,citecolor=blue]{hyperref}
\usepackage{color}
\usepackage{lipsum} 
\usepackage{nicefrac}
\usepackage{leftindex}
\usepackage{float}
\usepackage{subfigure}
\usepackage{dcolumn}
\usepackage{dsfont}
\usepackage{placeins} 
\usepackage{bm}

\newcommand{\im}{\mathrm{i}}
\newcommand{\e}{\mathrm{e}}
\begin{document}


\title{Resonant excitations via low frequency pumping in driven magnon systems}

\author{Jan~Mathis~Giesen}
	\email{jmgiesen@physik.uni-kl.de}
	\affiliation{Fachbereich Physik and Landesforschungszentrum OPTIMAS, Rheinland-Pf\"alzische Technische Universit\"at Kaiserslautern-Landau, 67663 Kaiserslautern, Germany}

\author{Alexandre~Abbass~Hamadeh}
	\email{alexandre.hamadeh@universite-paris-saclay.fr}
	\affiliation{ Université Paris-Saclay, Centre de Nanosciences et de Nanotechnologies, CNRS, 91120, Palaiseau, France}

\author{Imke~Schneider}
	\email{ischneider@physik.uni-kl.de}
	\affiliation{Fachbereich Physik and Landesforschungszentrum OPTIMAS, Rheinland-Pf\"alzische Technische Universit\"at Kaiserslautern-Landau, 67663 Kaiserslautern, Germany}

\author{Philipp~Pirro}
	\email{ppirro@rptu.de}
	\affiliation{Fachbereich Physik and Landesforschungszentrum OPTIMAS, Rheinland-Pf\"alzische Technische Universit\"at Kaiserslautern-Landau, 67663 Kaiserslautern, Germany}

 \author{Sebastian~Eggert}
	\email{eggert@physik.uni-kl.de}
	\affiliation{Fachbereich Physik and Landesforschungszentrum OPTIMAS, Rheinland-Pf\"alzische Technische Universit\"at Kaiserslautern-Landau, 67663 Kaiserslautern, Germany}


\date{\today}

\begin{abstract}
We analyse resonant excitations of ferromagnetic magnons via microwave pumping using Floquet theory.  Special focus is put 
on driving frequencies that are below the corresponding magnon energy, which can be excited in large parameter regions via
parametric resonances.  
We develop a theoretical framework that analytically predicts the regions of resonances and resonance thresholds in thin films of ferro- and ferri-magnetic materials like YIG as a function of damping, amplitude and frequency. Resonance regions are separated by exceptional points of the quasi-energies and the results are compared with micromagnetic simulations.  The corresponding threshold amplitudes can be estimated from a characteristic powerlaw with damping, leading  to 
the possibility of 
targeted exciatations at selected wavenumbers using low frequency drive.
\end{abstract}

\maketitle


{\it \label{sec:level1}Introduction.}
Parametric driving has emerged as a versatile tool across many areas of physics \cite{PhysRevA.92.023815,PhysRevLett.123.173601,PhysRevX.7.021015,PhysRevLett.115.205301,Geilen2025}. It can be used for amplification of signals \cite{PhysRevX.6.041026,Macklin2015,Malz2019} including spin waves \cite{BRACHER20171,10314026,lentfert2026}, to create squeezed states \cite{PhysRevResearch.6.033090,PhysRevLett.107.213603} and for the realization of magnon Bose-Einstein condensates \cite{Demokritov2006,10.1063/5.0189154,Dzyapko_2007,PhysRevLett.99.037205}. Furthermore, parametric driving can cause a sub-harmonic response of the system, making it a promising route toward the realization of discrete time crystals \cite{PhysRevA.110.L010202,PhysRevB.100.020406,PhysRevLett.126.243401,PhysRevLett.133.266601}. \\
Driving with an external magnetic field can create parametric excitations in ferro- or ferrimagnets like Yttrium-Iron-Garnet (YIG) \cite{Morgenthaler,SUHL1957209,Schloemann}. One of these processes is parallel pumping, where an oscillating magnetic field is applied parallel to a static bias field.  This method to excite magnons parametrically has been studied extensively in the past \cite{l2012wave,Rezende}. 
However, most research focuses on the first order resonance, where the system is driven with twice the frequency of the resonant mode. Much less research has been dedicated to resonances at lower driving frequencies despite the fact that higher-harmonic resonant magnon creation is also possible \cite{SAFONOV2019165486}. For example, higher order parametric resonances have already been observed in perpendicular driven Permalloy films \cite{Bauer2015}. Generally, higher order resonances enable amplification, squeezing and pumping at low driving frequencies, including below the lowest magnon energy band.

In this paper we establish an efficient method based on Floquet theory \cite{Floquet} to develop a microscopic understanding of these resonance processes which appear as instabilities in our system. Floquet theory is an established tool to deal with time periodic systems capable of finding resonances and steady states \cite{Reyes_2017,dauer25,giesen2025}, discovering new behaviors \cite{FloquetMagnons} and modifying systems through periodic driving i.e. Floquet engineering \cite{Holthaus_2016, Eckardt,Bukov04032015,PhysRevX.4.031027}.  To investigate the effect of damping on parametric resonances we additionally use a semi-classical ansatz based on the Landau-Lifshitz-Gilbert equation (LLG) \cite{Landau1935ONTT,Gilbert} which is used to generalize the Floquet equations. 
 Using degenerate perturbation theory we find analytic formulas for the thresholds of the first three orders of resonance providing a better understanding of the material parameter dependence of said thresholds. \\
Our main interest is investigating thin films of YIG. YIG plays an important role in experiments studying magnons and the development of magnonic devices due to its low damping \cite{CHEREPANOV199381,Kruglyak_2010,Serga_2010,PhysRev.110.73,PhysRevLett.100.257202}. We confirm our results by performing micromagnetic simulations for YIG. \\
Our studies show the existence of low frequency parametric resonances, the appearance of off-resonant Floquet replicas as well as a counter-intuitive vanishing of resonances at high driving amplitudes. We see that the introduction of Gilbert damping changes the size of instability regions.  The thresholds for parametric resonances show a characteristic power-law dependence, while the position of the instabilities are not affected.  The prediction of instability regions from Floquet theory are complimented by micromagnetic simulations in order to take into account magnon-magnon interactions.  \\

\begin{figure}
    \centering
    \includegraphics[width=1\linewidth]{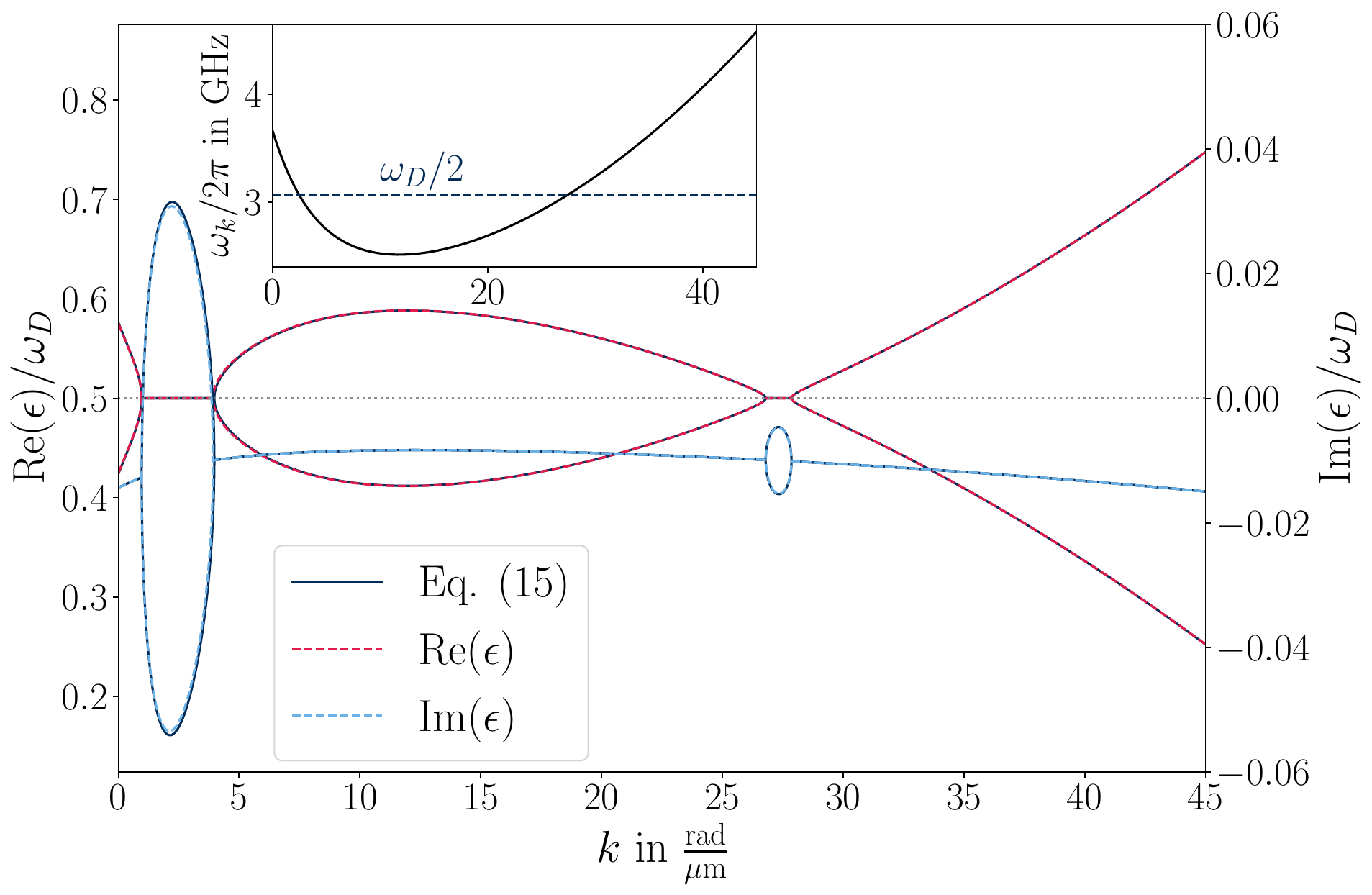}
    \caption{Analytical quasi-energies of Eq.~\eqref{eq:QE_n1_x} (solid) compared with numerical solution of Eq.~\eqref{eq:DGL2_damped} (dashed) for  $\omega_D \!=\! 6.13\,\mathrm{GHz}$,
    static field $h_0=700\,\mathrm{Oe}$, driving amplitude $h_1=400\,\mathrm{Oe}$, film thickness $d=0.5\mathrm{\mu m}$ and  damping $\alpha =0.02$. The inset shows the dispersion $\omega_k$ in Eq.~\eqref{eq:dispersion} with half the driving frequency plotted as a dashed line. \vspace{-.3cm}}
    \label{fig:QE_n1_pt}
\end{figure}

{\it \label{sec:level2}Theory.}
We examine a thin film of a ferrimagnetic material like YIG in periodic magnetic fields. 
The system is described by an effective  spin-$S$ Heisenberg  model with additional dipole-dipole interactions \cite{Kreisel2009}
\begin{align}
H=&-\frac{1}{2}\sum_{i\neq j}J_{i j}\mathbf{S}_i \cdot \mathbf{S}_j -h(t)\sum_iS_i^z \\
&+ \frac{1}{2}\sum_{i,j}\sum_{\alpha,\beta}D_{i j}^{\alpha\beta}S_i^\alpha S_j^\beta,
\end{align}
where $h(t)=h_0+h_1\cos(\omega_D t)$ is the energy of the magnetic field 
and the dipole-dipole tensor reads
\begin{align*}
D_{i j}^{\alpha\beta}=(1-\delta_{i j})\frac{4 \mu_B^2}{|\mathbf{r}_{i j}|^3}\left(\frac{3r_{i j}^\alpha r_{i j}^\beta}{|\mathbf{r}_{i j}|^2}-\delta^{\alpha\beta}\right),
\end{align*}
where $\mathbf{r}_{i j}\! =\! \mathbf{r}_j\!-\!\mathbf{r}_i$.  Working in units of twice the Bohr magneton $2\mu_B$ the parameters for YIG are $S\!=\!14.2$, $J/k_B\!=\! 2.74\mathrm{K}$, and lattice spacing $a\!=\!12.376 \mathrm{\r{A}}$ \cite{Serga_2010,PhysRev.110.73,PhysRevLett.100.257202}. \\
Using the Holstein-Primakoff transformation in the linear spin wave approximation \cite{PhysRev.58.1098}
\begin{align*}
S_i^+\approx \sqrt{2S}a_i,\qquad S_i^-\approx\sqrt{2S}a_i^\dagger ,\qquad S_i^z=S-a_i^\dagger a_i
\end{align*}
where terms beyond the leading order in $1/S$ are neglected,
we arrive at a Hamiltonian quadratic in bosonic creation and annihilation operators. We impose periodic boundary conditions and Fourier transform into momentum space. To describe the lowest magnon band, we furthermore employ the uniform-mode approximation \cite{Kreisel2009}, retaining only the $k_x=0$ mode in the thickness direction. We arrive at
\begin{align}
\label{eq:H}
H\approx &\sum_\mathbf{k} \left[A_\mathbf{k}(t) a_\mathbf{k}^\dagger a_\mathbf{k}+\frac{B_\mathbf{k}}{2}\left( a_\mathbf{k} a_{-\mathbf{k}} + a_\mathbf{k}^\dagger a_{-\mathbf{k}}^\dagger \right)\right] ,
\end{align} 
where $a_\mathbf{k}^\dagger$ ($a_\mathbf{k}$) are magnonic creation (annihilation) operators and the sum goes over all ${\mathbf k}=(k_y,\, k_z)^T$ in the plane of the film.   
An analytic expression for the full static dispersion is derived in \cite{Kreisel2009}, where the 
lowest energy occurs at $k_y\!=\!0$ and finite $k_z$ which is relevant for magnonic pumping \cite{Rezende}.  Accordingly, we want to consider the creation of excitations $k=k_z$ along the magnetic field direction, spatially averaged over the
other two directions $k_y\!=\!k_x\!=\!0$.  For this case, the coefficients are given by \cite{Kreisel2009}
\begin{align}
A_{k} &= h(t)+2JS(1\!-\!\cos(k a)) 
\!+\!\frac{8\pi S\mu_B^2}{a^3}\frac{1-\mathrm{e}^{-|k|d}}{|k|d}\\
B_k & = \frac{8\pi S\mu_B^2}{a^3}\frac{1-\mathrm{e}^{-|k|d}}{|k|d},
\end{align} 
where
$d$ is the thickness of the film which is assumed to be much smaller than the other dimensions of the film. The time averaged coefficient $\bar{A}_k$ corresponds to 
the static case ($h_1=0$) with dispersion relation \cite{Kreisel2009}
\begin{align}
\omega_{k} = \sqrt{\bar{A}_{k}^2-B_k^2} \enspace,
\label{eq:dispersion}
\end{align} 
plotted in Fig.~\ref{fig:QE_n1_pt}(inset).\\
The Heisenberg equations of motion for the driven model are obtained by taking the  expectation values of
the transverse spin components $S^{x,y}_k=\langle a_k \pm a_{-k}^\dagger\rangle$
\begin{align}
\frac{d}{dt}\mathbf{S}_k=
\mathrm{i}\left(\begin{array}{cc}
0 & -A_k+B_k\\
-A_k - B_k & 0
\end{array}\right)
\mathbf{S}_k
\label{eq:DGL2}
\end{align}
where $
   \mathbf{S}_k = (S_k^x, S_k^y )^T$. 
In this formulation, the phenomenological Gilbert damping $\alpha$ can also be included
\begin{align} 
\frac{d}{dt}\mathbf{S}_k
&=
\frac{1}{1+\alpha^2}
\begin{pmatrix}
-\alpha(A_k+B_k) & -\mathrm{i}(A_k-B_k)\\
-\mathrm{i}(A_k+B_k) & -\alpha(A_k-B_k)
\end{pmatrix}
\mathbf{S}_k.
\label{eq:DGL2_damped}
\end{align}
which we recognize as the linearized LLG
equation \cite{Kreisel2009,Rezende}.
The dimensionless factor $\alpha$  changes for different energy regimes and is generally determined phenomenologically. YIG in particular can be fabricated with very low damping $\alpha\! =\! 0.0002$ \cite{Hauser2016,PhysRevLett.122.197201,Mohseni_2020}.
Eq.~(\ref{eq:DGL2_damped}) 
has been used to describe the fundamental parametric resonance around $\omega_D\!\approx\! 2 \omega_k$ in the rotating-wave approximation \cite{Rezende,l2012wave,BRACHER20171}. 
Higher-order resonances, however, are not captured within this approximation. We therefore employ Floquet theory to analyze them. 
According to Floquet theory, solutions to Eq.~\eqref{eq:DGL2_damped} can be expressed in the form of Floquet states \cite{Floquet,Rand} 
\begin{align}
  \mathbf{S}_k = e^{-\im \epsilon t}\mathbf{p}(t)  
  \label{floquet_ansatz}
\end{align} where $\epsilon$ is the quasi-energy and $\mathbf{p}(t)=\mathbf{p}(t+T)$  is a periodic function with the same period as the drive, $T=2\pi/\omega_D$. The real part of the quasi-energy corresponds to the energy of the magnons (modulo the driving frequency) in the driven system. 
The sign of the imaginary part of the quasi-energy determines the stability of the Floquet solution: $\mathrm{Im}(\epsilon)\leq0$ corresponds to a stable (decaying or marginal) solution, while $\mathrm{Im}(\epsilon)>0$ signals an instability. \\
To calculate the quasi-energies we first find the eigenbasis of the 
static part of the linear differential Eq.~\eqref{eq:DGL2_damped}.  The solution are eigenmodes with $\mathbf{p}$ constant in  Eq.~(\ref{floquet_ansatz}) where the quasi-energies are given by the eigenvalues
\begin{align}
\epsilon^{(0)}=-\mathrm{i}\alpha \bar{A}_{k}\pm\omega_k. \label{e0}
\end{align}
Thus, in the absence of the drive ($h_1\to0$), both eigenmodes decay exponentially due to the finite damping. 
Transforming the equation of motions into the corresponding eigenbasis we get
\begin{align} \label{eq:DGL_diag} \frac{\mathrm{d}}{\mathrm{d}t}\tilde{\mathbf{S}}_k = \left( -\alpha \bar{A}_{k}\mathbf{1} -\mathrm{i}\omega_k\sigma_z +h_1\cos(\omega_D t)\tilde{A}_1 \right) \tilde{\mathbf{S}}_k , \end{align}
where $\mathbf{1}$ denotes the identity matrix, $\sigma_z$ is the Pauli matrix, and $\tilde{A}_1$ is to linear order in $\alpha$
\begin{align*} \tilde{A}_1 = \mathrm{i}\left(\begin{array}{cc} \mathrm{i}\alpha - \frac{\bar{A}_{k}}{\omega_k} & -\frac{\Bar{A}_{k}}{\omega_k}+\frac{\mathrm{i}\alpha B_k + \omega_k}{\bar{A}_{k}+B_k}\\ \frac{\bar{A}_{k}}{\omega_k}+\frac{\mathrm{i}\alpha B_k - \omega_k}{\bar{A}_{k}+B_k} & \mathrm{i}\alpha + \frac{\bar{A}_{k}}{\omega_k} \end{array}\right). \end{align*}

Using the periodicity in  Eq.~\eqref{floquet_ansatz} we Fourier expand the solutions 
\begin{align}
\tilde{S}_k^\sigma = \mathrm{e}^{-\mathrm{i}\epsilon t}\sum_n \mathrm{e}^{-\mathrm{i}m\omega_D t} |\sigma,m\rangle
\end{align}
where $\sigma=\tilde x,\tilde y$ and $\epsilon$ is the (complex) quasi-energy. Inserting this in Eq.~\eqref{eq:DGL_diag} we obtain an algebraic Floquet equation
\begin{align}
\nonumber
&\left[(-\alpha \bar{A}_{k} + \mathrm{i} m\omega_D +\epsilon)\delta_{\sigma \sigma'} - \mathrm{i}\omega_k (\sigma_{z})_{\sigma\sigma'} \right]|\sigma',m\rangle\\
&+\frac{h_1}{2} (\tilde{A}_1)_{\sigma\sigma'}\left(|\sigma',m+1\rangle+|\sigma',m-1\rangle\right)=0,
\label{eq:AlgFE}
\end{align}
where the sum over $\sigma'$ is implicit.  This linear equation can be solved analytically or numerically to determine the Floquet solutions and the eigenvalues $\epsilon$.\\
Analytically, we apply degenerate perturbation theory in Floquet space, treating the time-dependent drive $h_1/\omega_D$ as a small parameter. 
For $\nicefrac{h_1}{\omega_D} \ll 1$ we treat the off-diagonal term on the second line in Eq.~\eqref{eq:AlgFE}  perturbatively. 
The unperturbed matrix becomes degenerate for the states $|\tilde x,m\rangle^{(0)}$ and $|\tilde y,m-n\rangle^{(0)}$ when
\begin{align}
n\omega_D=2\omega_k,
\label{eq:resonanceCond}
\end{align}
recovering the known resonance condition for parametric resonances. In the following we calculate the quasi energies close to the degeneracies but including small deviations $\Delta\omega_n=\omega_D-2\omega_k/n$. We first focus on the fundamental resonance $n=1$.  
Diagonalizing Eq.~\eqref{eq:AlgFE} within the degenerate subspace to first order in $h_1/\omega_D$ yields  
\begin{align}
\epsilon \approx& \omega_k -\im \alpha \bar{A}_{k} +\frac{1}{2}\Delta\omega_1 
\pm\im\sqrt{-\frac{\Delta\omega_1^2}{4}+\frac{h_1^2 B_k^2}{4\omega_k^2}},
\label{eq:QE_n1_x}
\end{align}
This result agrees very well with the exact numerical evaluation of Eq.~(\ref{eq:AlgFE}), even for relatively strong driving amplitudes $h_1/2\mu_B\!=\!400\,\mathrm{Oe}$ and $h_0/2\mu_B\!=\!700\,\mathrm{Oe}$ in Fig.~\ref{fig:QE_n1_pt}. Here, the
driving amplitude $\omega_D \!=\! 6.13\,\mathrm{GHz}$ 
was chosen not too far from the dispersion  minimum, where the resonance condition in Eq.~(\ref{eq:resonanceCond}) for $n=1$ 
is fulfilled for {\it two} values $k\!=\!{2.5 \rm rad\over \mu m}$  and ${27.4 \rm rad\over \mu m}$ (see inset).
 A larger value of damping $\alpha=0.02$ has been chosen for better visibility. 
Away from the resonances  Re$(\epsilon)\!\sim\!\omega_k, \,\omega_D\!-\!\omega_k$ follows the dispersion from the inset, but  
near resonant $k$ values 
the real parts become exactly degenerate and the imaginary parts split, leading to regions with {\it gain} Im$(\epsilon)\!>\!0$  around $k\!=\!2.5\frac{\mathrm{rad}}{\mathrm{\mu m}}$.  
According to Eq.~(\ref{floquet_ansatz})  this corresponds to exponentially growing excitations, i.e.~resonant instabilities. The resonances are linked to the off-diagonal terms in the dipole coupling, which is also the reason why the resonance is stronger (larger Im$(\epsilon)$) for smaller $k$ values, where long range effects dominate.

Notably, there are points where both  real and  imaginary parts of the quasi-energies are degenerate, i.e.~when the square root in Eq.~\eqref{eq:QE_n1_x} becomes zero. These so-called exceptional points are known to yield non-trivial topological properties in non-Hermitian systems \cite{Heiss_2012,RevModPhys.93.015005,ll76-j2l5,Downing2023,sidorenko2026} .

\begin{figure}[t]
\centering
\includegraphics[width=0.45\textwidth ]{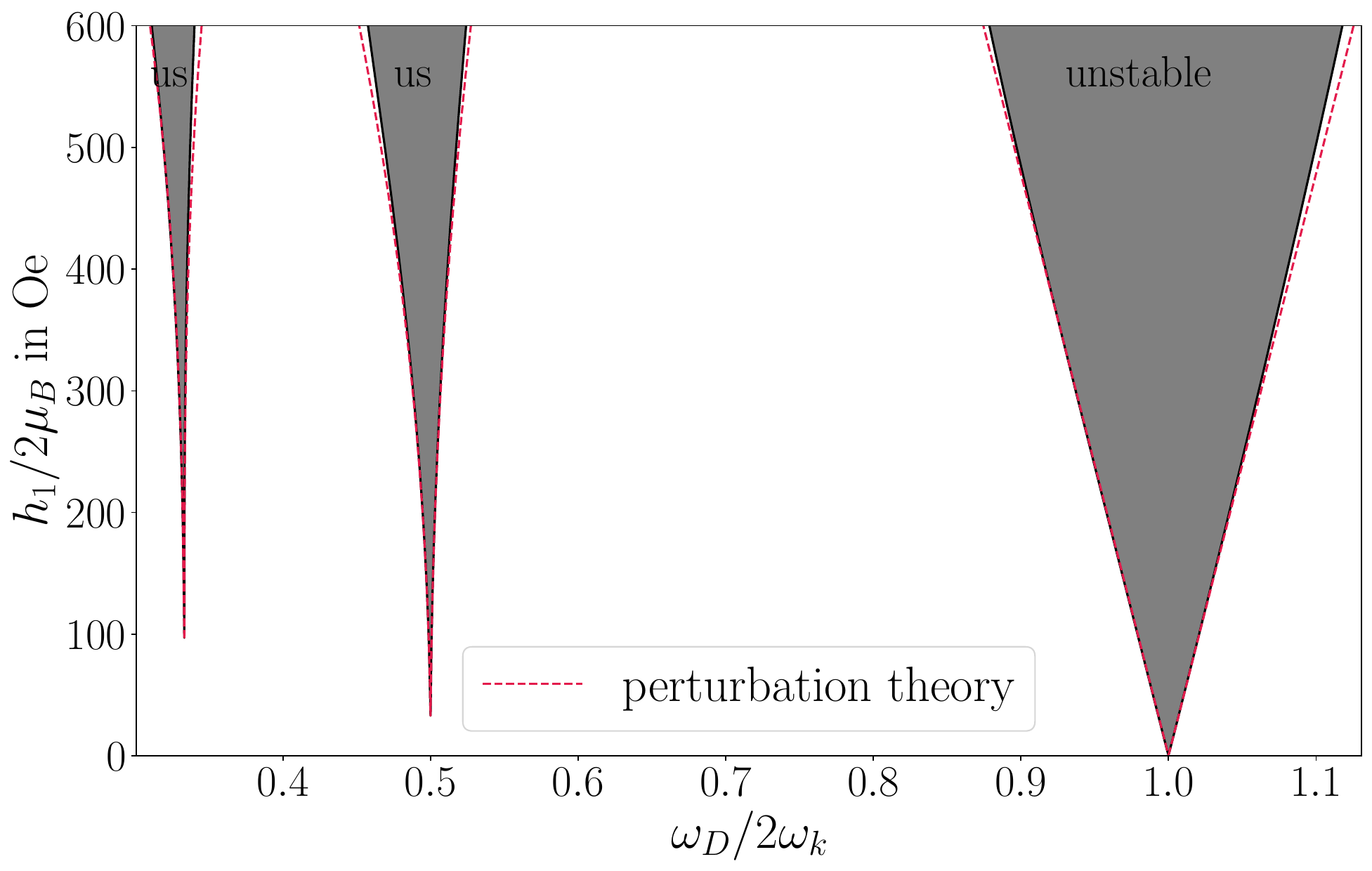}
\caption{Stability chart for excitations at $k\!=\!2.5\frac{\mathrm{rad}} {\mathrm{\mu m}}$, static field $h_0/2\mu_B\!=\!700\,\mathrm{Oe}$, film thickness $d\!=\!400a$ and damping $\alpha\!=\!0.0002$. The grey areas show the parameter regions where 
Eq.~\eqref{eq:DGL2_damped} has unstable solutions Im$(\epsilon)\!>\!0$. The red dashed lines are the analytical predictions for the thresholds in Eqs.~\eqref{eq:hthn1}, \eqref{eq:hthn2} and \eqref{eq:hthn3}, the black solid lines the numeric ones. \vspace{-.3cm} }
\label{fig:stability}
\end{figure}
To  visualize the resonant instabilities in Fig.~\ref{fig:stability} we have numerically identified regions where Im$(\epsilon)\!>\!0$ for excitations at fixed $k\!=\!{2.5 \rm rad\over \mu m}$ as a function of $\omega_D$  at low damping $\alpha\!=\!0.0002$.  We observe resonances near $\omega_D=2\omega_k/n$ for $n=1,2,3$ in Eq.~(\ref{eq:resonanceCond}).  For larger $n$ the instability regions are
thinner.  In the following we seek to analytically determine the threshold driving amplitudes $h_{\rm th}$,  defined  by the condition $\mathrm{Im}(\epsilon)\!=\!0$.

For $n=1$ we use Eq.~(\ref{eq:QE_n1_x}) to obtain
\begin{align}\label{eq:hthn1}
h_\mathrm{th}&= \sqrt{h_\mathrm{th,min}^2+\frac{\omega_k^2}{B_k^2}\Delta\omega_1^2}, \ \ \ 
h_\mathrm{th,min} &=2\alpha\omega_k\frac{\bar{A}_{k}}{B_k},
\end{align}
corresponding to a region with very small on-resonance threshold proportional to $\alpha$ for $\Delta\omega_1=0$, which expands linearly for larger $h_{\rm th} \propto \Delta \omega_1$.

To calculate the threshold of the $n=2$ instability at $\omega_D=\omega_k+\Delta\omega_2$, the degeneracy is lifted in second order in $h_1$, yielding
\begin{align}
\epsilon \!\approx& \omega_k \!-\!\mathrm{i}\alpha A_k\! +\!\Delta\omega_2 
\pm\mathrm{i}\sqrt{\!-\!\left(\Delta\omega_2 \!-\!c_0h_1^2\right)^2+c_1h_1^4},
\end{align}
where $c_0 = -\frac{B_k^2}{3\omega_k^2}$ and $c_1=\frac{\bar{A}_k^2B_k^2}{4\omega_k^6}$.
Working to leading order in $\Delta\omega_2$ and $\alpha^2$ terms, 
we obtain
\begin{align}\label{eq:hthn2}
h_\mathrm{th}&\approx\sqrt{\frac{c_0\Delta\omega_2}{c_1-c_0^2}+\sqrt{\frac{c_1\Delta\omega_2^2}{(c_1-c_0^2)^2}+\frac{\alpha^2\bar{A}_k^2}{c_1-c_0^2}}} \\
h_\mathrm{th,min}&\approx \omega_k\sqrt{\frac{2\alpha\omega_k}{B_k}}.
\end{align}
Analogously, 
the third order instability at $\omega_D \!=\!\frac{2}{3}\omega_k \!+\!\Delta\omega_3$ is obtained by including third order corrections $h_1^3$
\begin{align}
\nonumber
\epsilon=&\omega_k -\mathrm{i}\alpha A_k +\frac{3}{2}\Delta\omega_3 \\
&\pm\mathrm{i}\sqrt{-\left(\pm\Delta\epsilon^{(2)}-\frac{3}{2}\Delta\omega_3\right)^2 +\tilde{c}_1}
\label{eq:hthn3}
\end{align} 
\begin{align}
{\rm where \ \ \ }\Delta\epsilon^{(2)} = \leftindex^{(0)}\langle x,0|V_1|x,0\rangle^{(1)}\approx -\mathrm{i}h_1^2\frac{9B_k}{32\omega_k^3} & 
\end{align}
is the quasi-energy correction in second order with $|x,0\rangle^{(m)}$ the $m$th order correction to the eigenstate and
\begin{align*}
\tilde{c}_1=\leftindex^{(0)}\langle y,3|V_1|x,0\rangle^{(2)}\leftindex^{(0)}\langle x,0|V_1|y,3\rangle^{(2)}.
\end{align*} 
The minimal threshold at $\Delta\omega_3 =\Delta\epsilon_{x,0}^{(2)}$ is
\begin{align}
h_\mathrm{th,min}=\omega_k\left(\frac{2\alpha \bar{A}_k}{B_k\sqrt{c_k}}\right)^\frac{1}{3}
\end{align}
\begin{align*}
{\rm where \ \ \ }c_k \approx \frac{81}{64\omega_k^4}\left(\bar{A}_{k}^4+\frac{1}{4}\bar{A}_k^2B_k^2 + \frac{1}{64}B_k^4\right).
\end{align*}
We can also calculate the width of the instability depending on the driving amplitude $h_1\geq h_\mathrm{th,min}$
\begin{align}\label{eq:hthn3}
\Delta\omega_3 = \frac{2}{3}\Delta\epsilon_{x,0}^{(2)} \pm \frac{2}{3}\alpha \bar{A}_{k}\sqrt{\left(\frac{h_1}{h_\mathrm{th,min}}\right)^6-1}.
\end{align}

We find good agreement with the analytically calculated threshold for the first three orders of instability in Eqs.~\eqref{eq:hthn1}, \eqref{eq:hthn2} and \eqref{eq:hthn3}, shown as red lines in Fig.~\ref{fig:stability} with slight deviations for larger $h_1$ and $\Delta\omega_n$.  The case of very strong driving is considered below.

\begin{figure}[t]
\centering
\includegraphics[width=0.4\textwidth]{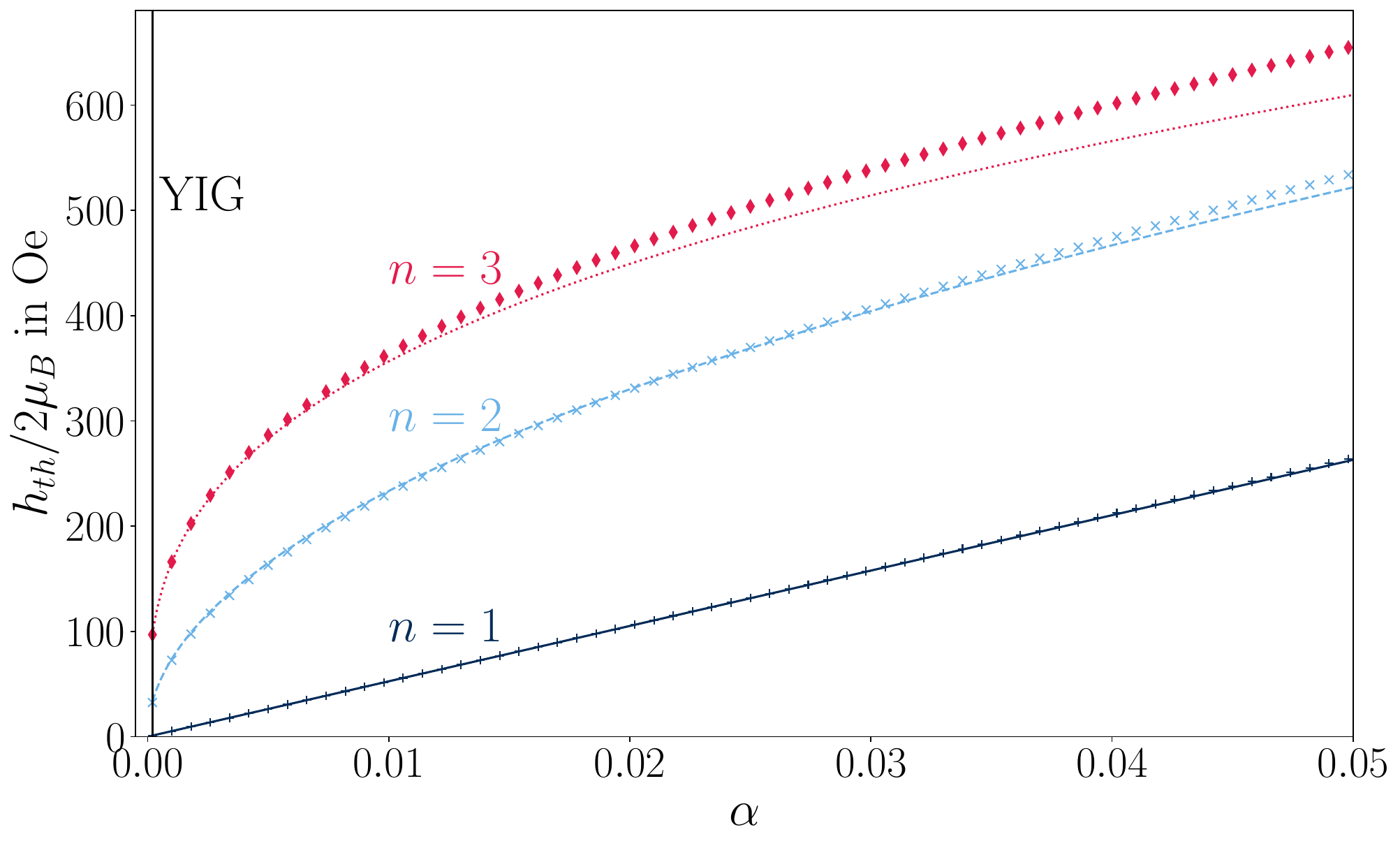}
\caption{The behavior of the minimal threshold for the mode $k\!=\!2.5\frac{\mathrm{rad}}{\mathrm{\mu m}}$ for different dampings $\alpha$ and orders of resonance $n$ at $h_0\!=\!700\mathrm{Oe}$ and and film thickness $d\!=\!400a$. The markers are the full numerical results and the connected lines the analytical predictions. The black vertical line marks the assumed damping of YIG. }\vspace{-.3cm}
\label{fig:guess}
\end{figure}

In general higher order resonances $n\omega_D = 2 \omega_k$ correspond to degeneracies with the $n^{\rm th}$ Floquet replica,  which in turn require powers of $h_1^n$ in perturbation theory to lift 
the degeneracy.  We therefore predict that 
\begin{align}
h_\mathrm{th,min} \propto \alpha^\frac{1}{n}, 
\end{align}
which is confirmed numerically in Fig.~\ref{fig:guess} as a function of $\alpha$ for $n=1,2,3$.  For larger thresholds outside the perturbative limit deviations can be observed.

\begin{figure}[t]
\centering
\includegraphics[width=0.44\textwidth]{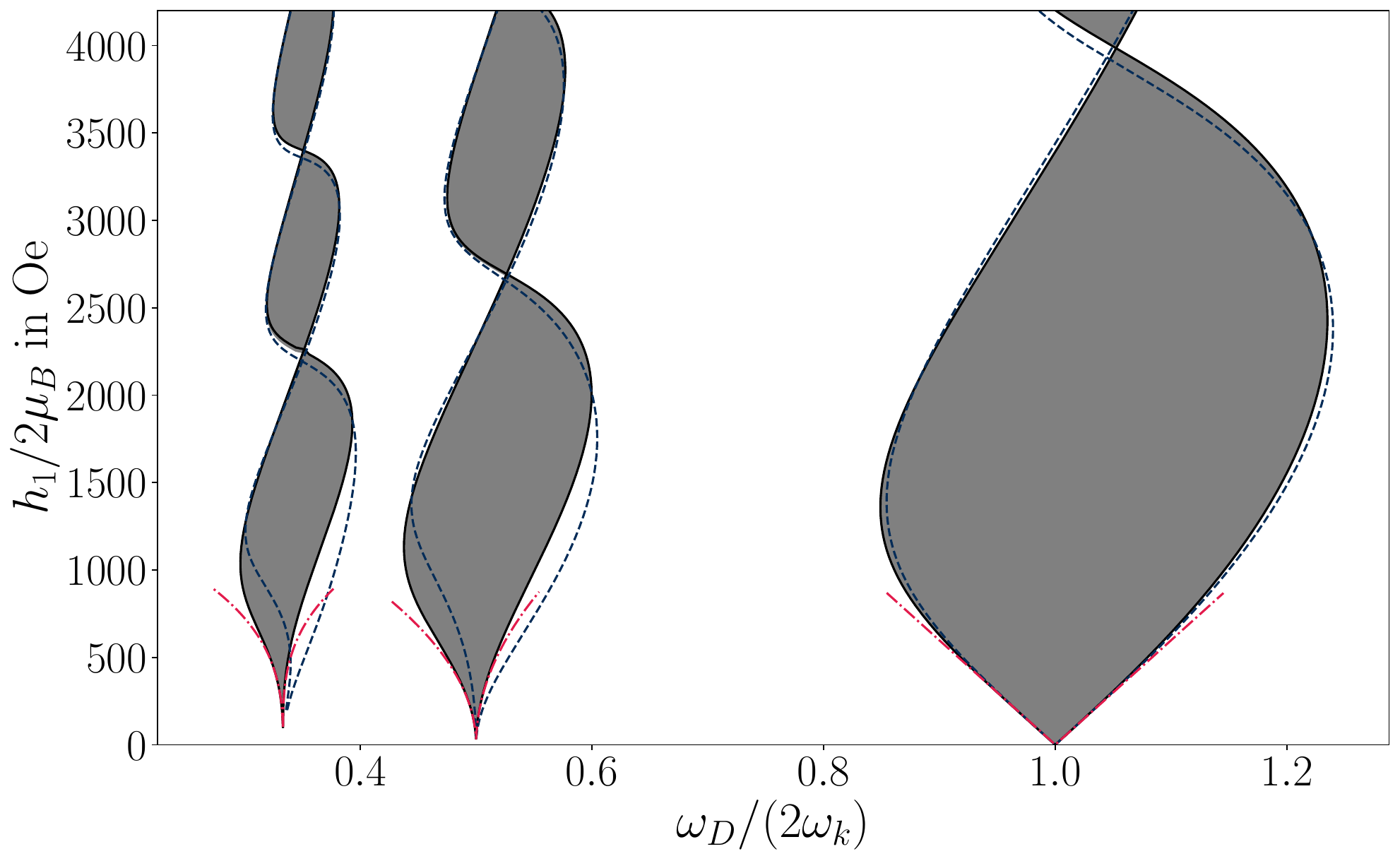}
\caption{Stability regions for $k=5\,\frac{\mathrm{rad}}{\mathrm{\mu m}}$, $h_0=700\,$Oe, $\alpha=0.0002$ and and film thickness $d=400a$ as a function of driving frequency $\omega_D$ and driving amplitude $h_1$. The blue dashed lines show the analytical result using Eq.~\eqref{eq:instability_tr} and the red dashed-dotted line is the weak driving approximation in Eqs.~\eqref{eq:hthn1}, \eqref{eq:hthn2} and \eqref{eq:hthn3}. \vspace{-.3cm}}
\label{fig:Stability_fh}
\end{figure}

We now consider larger driving $h_1$ which shows  
additional features  not captured by perturbation theory in $h_1$.  
Progress can be made for $\frac{\omega_D}{h_1} \!\gg\! 1$
by performing a gauge transformation 
$Q(t) \!=\! \exp[-\mathrm{i}\frac{h_1}{\omega_D}\sin(\omega_D t)(a_k^\dagger a_k+a_{-k}^\dagger a_{-k})]$ to the Schrödinger equation  \cite{Eckardt_2015} 
\begin{align}
    Q^\dagger (H-\im\partial_t)Q=\tilde{H}-\im \partial_t.
\end{align}
The Hamiltonian in Eq.~\eqref{eq:H} is transformed to 
\begin{align}
\label{eq:Hg}
\tilde{H}&=\sum_k\left( \bar{A}_k a_k^\dagger a_k + \left(\frac{\tilde{B}_k(t)}{2}a_k^\dagger a_{-k}^\dagger + h.c.\right)\right)\\
\tilde{B}_k &= B_k\sum_{n=-\infty}^{\infty} \mathcal{J}_n\!\left(\frac{2h_1}{ \omega_D}\right)\text{e}^{\text{i}n\omega_D t}
\end{align}
where $\mathcal{J}_n$ is the $n$th Bessel function.
The transformation transfers the time dependence from the diagonal term in the Hamiltonian to the off-diagonal term. 
Further, the time dependent term $\tilde{B}_k(t)$ is no longer proportional to $h_1$ but now 
depends on $B_k\mathcal{J}_n\!\left(\frac{2h_1}{\omega_D}\right)$. If this is small compared to the driving frequency we can again apply degenerate perturbation theory, though it now stays valid even for large $h_1$. In appendix \ref{app:highFields} we derive an approximate quasi-energy
\begin{align}
  \epsilon\! =\! \tilde{\omega}_k\! +\!\frac{n}{2}\Delta\omega_n+\im \sqrt{\mathcal{J}_n^2\!\left(\frac{2h_1}{\omega_D}\right)B_k^2\!-\!\left(\frac{n}{2}\Delta\omega_n\right)^2},
\label{eq:qe_transformed}
\end{align}
\begin{align}
 {\rm where \ \ \ }   \tilde{w}_k = \sqrt{\bar{A}_k^2-\mathcal{J}_0^2\!\left(\frac{2h_1}{\omega_D}\right)B_k^2}
    \label{eq:omega_eff}
\end{align}
is the eigenfrequency of the time independent part of the Hamiltonian $\tilde{H}$. 
Positive arguments of the square root in Eq.~\eqref{eq:qe_transformed}  lead to a resonance Im$(\epsilon)>0$. This defines the border of the instability region 
\begin{align}
\frac{n}{2}\omega_D = \tilde{\omega}_k \pm  \mathcal{J}_n\!\left(\frac{2h_1}{\omega_D}\right)B_k.
\label{eq:instability_tr}
\end{align}
According to Eq.~\eqref{eq:instability_tr} the width of the $n$th instability region is proportional to the $n$th Bessel function. Therefore we expect the width to decrease with large $h_1$ and to vanish 
at the zeros of the Bessel functions, which agrees with numerical results 
in Fig.~\ref{fig:Stability_fh}.

\noindent
Generally,  Eq.~\eqref{eq:instability_tr} describes this behavior very well, especially for very large $h_1$, while 
the prediction from Eqs.~\eqref{eq:hthn1}, \eqref{eq:hthn2} and \eqref{eq:hthn3} (red dashed-dotted lines) fit better for small  and intermediate $h_1$.
We also see that at  high driving fields the instabilities get shifted to higher frequencies, which can be explained by a change of the magnon dispersion \eqref{eq:omega_eff} and therefore our resonance condition.  The reduction at the zeros of the Bessel function only occurs at the selected wavenumber and field, while other
$k$-values are still exponentially enhanced by the strong drive.   Therefore, the high amplitude effect cannot be observed as an overall reduction of excitations in realistic 
simulations or experiments.

{\it \label{sec:level4}Simulations.}
To further investigate the periodically driven system at low frequencies, we will look at the example of driving with $\omega_D=\nicefrac{2}{3}\,\omega_k$ for a selected mode. The goal is to validate the results of the previous section by comparing them to micromagnetic simulations. We use the open source software MuMax3 \cite{mumax} to numerically solve the complete LLG-equation
\begin{align}
\frac{\text{d}}{\text{d}t}\vec{S}_i \!=\! \frac{\gamma}{(1+\alpha^2)}\vec{S}_i\!\times\! \vec{H}_i\!-\!\frac{\alpha \gamma}{S(1\!+\!\alpha^2)}\vec{S}_i\!\times\!(\vec{S}_i\!\times\!\vec{H}_i)
\label{eq:LLG}
\end{align}
where $\vec{H}_i$ is the effective field acting on the magnetisation $\vec{S}_i$ at site $i$ and $\gamma$ the gyromagnetic ratio.\\
To get the magnon energy spectrum we have to Fourier transform the solution $S(\vec{x},t)$ in space and in time. As discussed in the previous analysis, for weak driving we generally expect solutions oscillating with $\omega \approx \omega_{k,0} + m\omega_D$, $m\in \mathbb{Z}$. Compared to the static case, this leads to copies of the dispersion, so called Floquet replicas.  Further, if the driving field $h_1$ exceeds the threshold, we expect the resonant modes in the instability region to show an exponential larger magnon density. Though, unlike in our linear approximation \eqref{eq:DGL2_damped}, we expect it to reach a saturation in the simulation for long enough simulation times due to higher order processes like magnon magnon scattering. We simulate a system which is $180\,\mathrm{\mu m}$ in $z$-direction, which is the direction of the static magnetic field $h_0$, a film thickness of $0.5\,\mathrm{\mu m}$ in $x$-direction and with periodic boundary conditions in $y$-direction. A finite size in $y$-direction is also possible, though it will affect the shape of the magnon dispersion and position of its minimum \cite{Mohseni_2020}. We specifically look at magnons propagating in $z$-direction. At the borders we exponentially increase the damping to avoid reflections there. 
\begin{figure*}[htp]
\centering
\subfigure[$h_1/2\mu_B=160\,$Oe]{\includegraphics[width=0.32\textwidth]{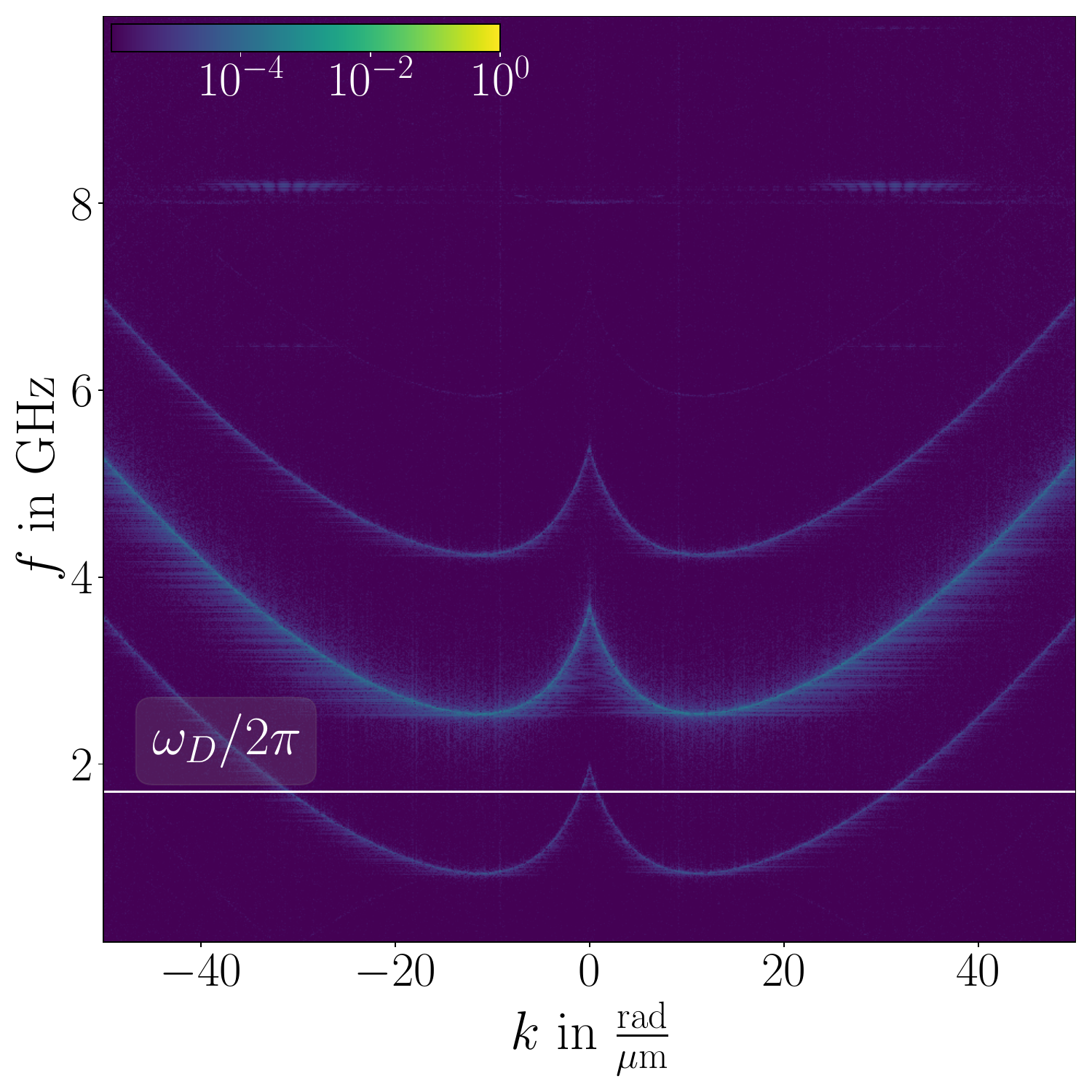}\label{fig:mumax_n3_a}}
\subfigure[$h_1/2\mu_B=190\,$Oe]{\includegraphics[width=0.32\textwidth]{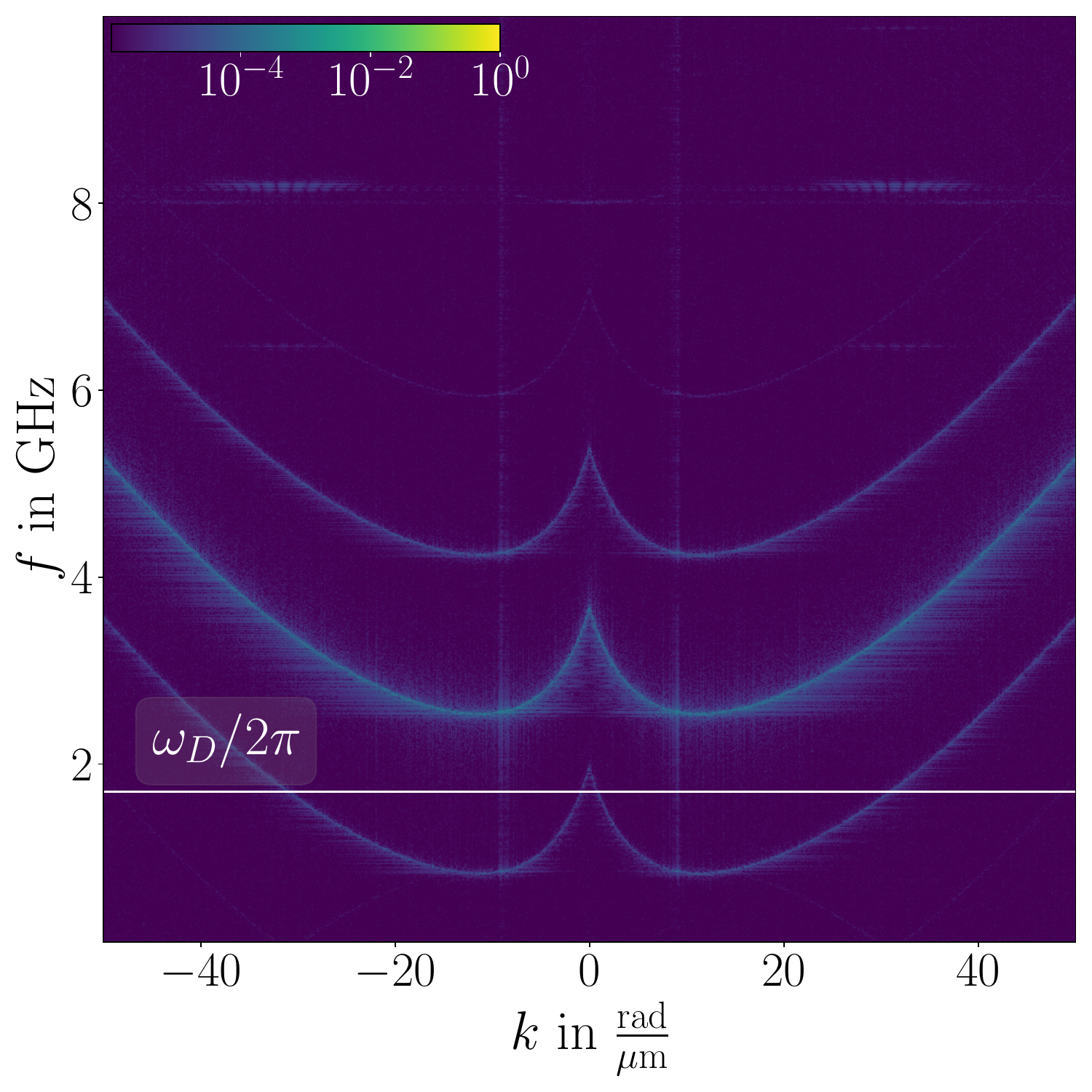}\label{fig:mumax_n3_b}}
\subfigure[$h_1/2\mu_B=300\,$Oe]{\includegraphics[width=0.32\textwidth]{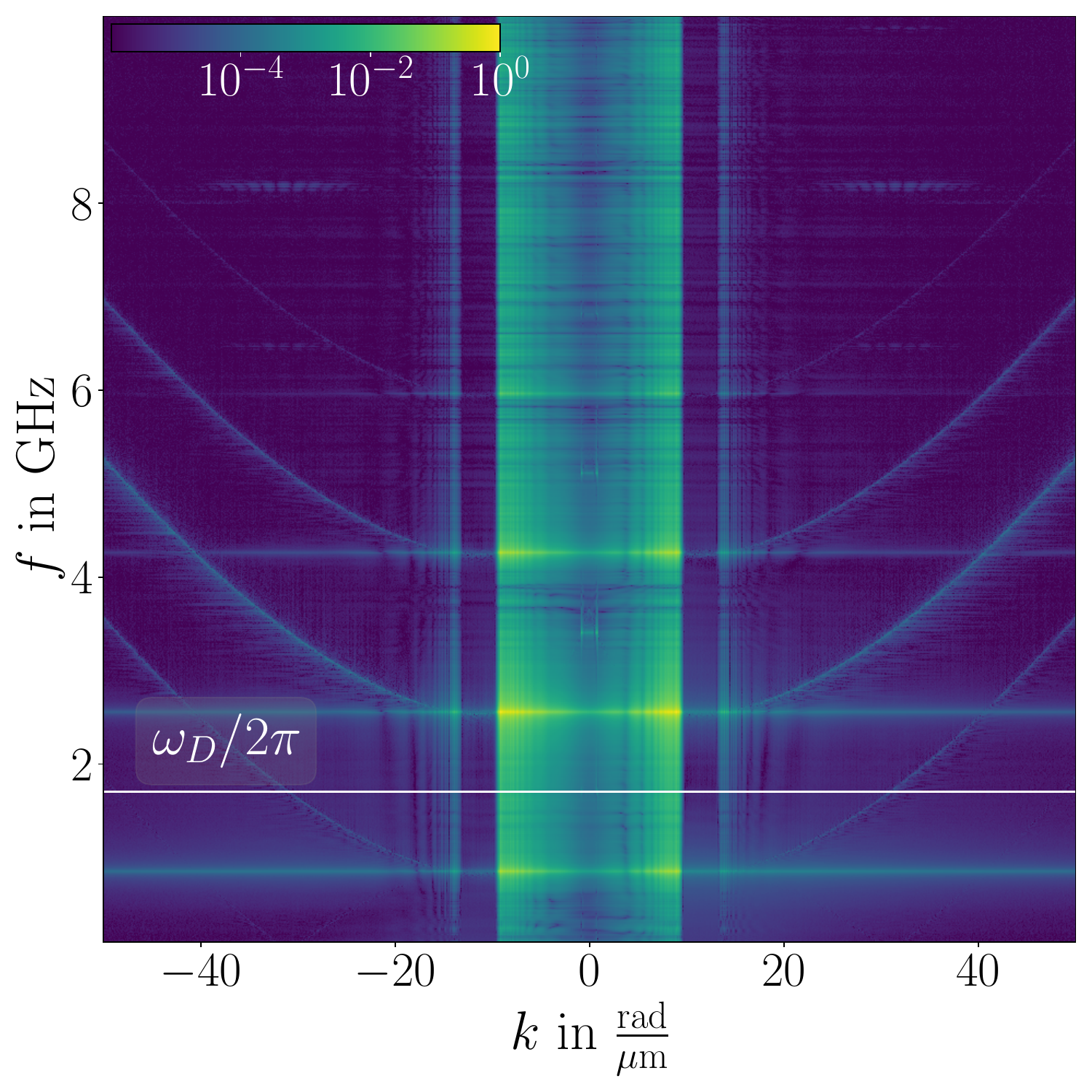}\label{fig:mumax_n3_c}}
\caption{ Colormap of the simulated magnon spectrum of YIG under periodic drive at the $n=3$ resonance $\frac{\omega_D}{2 \pi}=1.703\,\mathrm{GHz}$ with different driving amplitudes for a simulation time of $170\,\mathrm{ns}$. The color corresponds to the logarithmic intensity of the Fourier spectrum, normalized by the maximum Fourier amplitude, which occurs for 300Oe.}
\label{fig:mumax_n3}
\end{figure*}
\noindent
Results of such simulations are depicted in Fig. \ref{fig:mumax_n3}. Below the threshold (Fig. \ref{fig:mumax_n3_a}) we find the magnon dispersion as expected plus the Floquet replicas in a distance of $\nicefrac{\omega_D}{2\pi}$ around the original dispersion. These are the decaying solutions of the driven system, hence no $k$-value is significantly increased. The Floquet replicas are also smaller in intensity than the main dispersion. Furthermore, the intensity is rapidly decreasing for higher orders such that a second replica is only barely visible. Slightly above the threshold (Fig. \ref{fig:mumax_n3_b}) we see an increase in the magnon density for modes around $k\sim 8.6\nicefrac{\mathrm{rad}}{\mathrm{\mu m}}$ as predicted in the main dispersion as well as in the Floquet replicas. If we increase the driving (Fig. \ref{fig:mumax_n3_c}), so does the number of magnons at these modes due to the non-linear behaviour, becoming much larger than for non-resonant modes. We still find Floquet replicas in a distance $m\nicefrac{\omega_D}{2\pi}$ of the main excitation. We also see a weaker excitation of the mode around $k\sim 15.5\nicefrac{\mathrm{rad}}{\mathrm{\mu m}}$, which has the same frequency. That we see a higher threshold and weaker response to the drive at higher frequencies is consistent with the result of our perturbative analysis. The parametric resonance depends on the dipole-dipole interaction, which, while important for large wavelengths, becomes less significant for larger $k$ compared to the exchange interaction. We note that the thresholds observed in the micromagnetic simulation is slightly higher than the ones calculated using the linearized equations. 
 Higher effective damping may be due to the
finite size system with increased damping at the borders as well as due to magnon-magnon scattering, which was neglected in the analytic calculations.

{\it \label{sec:level4}Conclusion.}
We examine parametric resonances in thin films of YIG driven by an oscillating magnetic field. 
Using Floquet theory, we calculate the system's quasi-energies analytically and numerically. Notably, parametric resonances are directly tied to degeneracies in the real quasi-energy spectrum, with exceptional points separating regimes of fundamentally different physical behavior.  

We identify and map multiple orders of resonance with good agreement between numerics and analytics in the relevant parameter regimes. Analytic predictions for resonance thresholds are derived using degenerate perturbation theory in Floquet space. We find a power-law dependence between the thresholds and the Gilbert damping.
At very high driving amplitudes 
Bessel functions govern the instability regions. 

It should be noted that  higher order resonances at lower frequency driving  are  much less studied.  We showed that such pumping generally requires larger threshold
values due to the powerlaw behavior with damping.  
The typical threshold amplitude for the $n=3$ instability was predicted to be around 200Oe, which remains realistic and would allow resonant creation of 
excitations at a driving frequency  $\omega_D=2 \omega_k/3 < \omega_k$ below the energy of the magnon.  Therefore, linear excitations of magnons and higher order resonances are suppressed
at this frequency, which opens the door to a highly targeted creation of magnons at the selected wavenumber.  This has been confirmed by realistic micromagnetic simulations.


\bibliography{bibliography.bib}
\appendix

\section{Calculation of vanishing resonances}
\label{app:highFields}
The goal is to calculate the quasi-energies for large driving amplitudes $h_1$ in a perturbative way. Diagonalizing the static part of the gauge transformed Hamiltonian~\eqref{eq:Hg} yields with $\tilde{H}=\sum_n \tilde{H}_k$
\begin{align*}
\tilde{H}_k =& \left(\tilde{\omega}_{k}-B_k \sum_{n\neq 0}c_{2n}(t)\frac{B_k}{\tilde{\omega}_k} \right)(b_k^\dagger b_k+b_{-k}b_{-k}^\dagger)\\
&+B_k \sum_{n\neq 0}\left( c_{2n-1}(t)(b_k^\dagger b_{-k}^\dagger - b_kb_{-k})\right. \\
& \qquad\qquad\left. +c_{2n}(t)\frac{A_k}{\tilde{\omega}_k}(b_k^\dagger b_{-k}^\dagger + b_kb_{-k})\right)
\end{align*}
where $\tilde{\omega}_k$ is given by Eq.~\eqref{eq:omega_eff}, $c_n(t)=\mathcal{J}_{n}\left(\frac{2h_1}{\omega_D}\right)\mathrm{e}^{\mathrm{i}n\omega_D t}$, $b_k = u_ka_k+v_ka_{-k}^\dagger$, $u_k=\sqrt{\frac{A_k+\tilde{\omega}_k}{\tilde{\omega}_k}}$ and $v_k=\sqrt{\frac{A_k-\tilde{\omega}_k}{\tilde{\omega}_k}}$. Using this version of the Hamiltonian, we calculate the Heisenberg equations of motion
\begin{align*}
\mathrm{i}\dot{b}_k=&\tilde{\omega}_{k}b_k + B_k \sum_{n\neq 0}\left(c_{2n-1}(t)b_{-k}^\dagger \right.\\
&\qquad\qquad \left.+c_{2n}(t)\left(\frac{A_k}{\tilde{\omega}_k} b_{-k}^\dagger - \frac{B_k}{\tilde{\omega}_k} b_k\right)\right)\\
\mathrm{i}\dot{b}_{-k}^\dagger=&-\tilde{\omega}_{k}b_{-k}^\dagger + B_k \sum_{n\neq 0}\left(c_{2n-1}(t)b_{k}\right. \\
&\qquad\qquad \left.-c_{2n}(t)\left(\frac{A_k}{\tilde{\omega}_k} b_{k} - \frac{B_k}{\tilde{\omega}_k} b_{-k}^\dagger\right)\right).
\end{align*}
Analogous to the weakly driven case, we can go to Fourier space by using the spectral decomposition $b_k=\sum_m\e^{-\im(\epsilon+m\omega_D)t}b_k^{(m)}$. Then we perform degenerate perturbation theory in the degenerate subspace around $m\omega_D=2\tilde{\omega}_{k}$. Contrary to before, the leading order is now the first for all orders of resonance $n$. To obtain the quasi-energies we have to solve for odd $n$
\begin{align}
\mathrm{det}\left(\begin{array}{cc}
 \epsilon-\tilde{\omega}_k & \mathcal{J}_{n}\left(\frac{2h_1}{\omega_D}\right)B_k \\ -\mathcal{J}_{n}\left(\frac{2h_1}{\omega_D}\right)B_k & \epsilon -\tilde{\omega}_k- n\Delta\omega_n
\end{array}\right)=0,
\end{align}
where $\Delta\omega_n = \omega_D - \frac{2}{n}\tilde{\omega}_k $. For even $n$ we get a similar equation. 
This yields for both even and odd $m$ Eq.~\eqref{eq:qe_transformed}.

\section{Landau-Lifshitz-Gilbert equation}
Another way of approaching the problem of spin waves in periodically modulated fields is a semi-classical ansatz by solving the phenomenological Landau-Lifshitz (LL) eq.
\begin{align}
\frac{\text{d}}{\text{d}t}\vec{S}_i = \gamma\vec{S}_i\times (\vec{H}_\mathrm{ext}+\vec{H}_{\mathrm{eff},\,i})
\end{align}
where $\vec{H}_{\mathrm{eff},\,i}=-\frac{1}{\mu_0}\frac{\delta\epsilon}{\delta S_i}$, with $\epsilon$ the energy density, is the effective field acting on spin site $i$ stemming from exchange interaction and dipole-dipole interaction. We can perform a linear approximation by assuming that the spins only deviate little from their equilibrium position $S^x,S^y \ll S^z=S$. As a result for $k\neq 0$ the equations of motion take the form of a linear differential eq.
\begin{align}
\frac{\text{d}}{\text{d}t}\left(\begin{array}{c}
S_k^x \\ S_k^y
\end{array}\right)=
\left(\begin{array}{cc}
0 & A_k-B_k\\
-A_k - B_k & 0
\end{array}\right)
\left(\begin{array}{c}
S_k^x \\ S_k^y
\end{array}\right)
\label{eq:LL_lin}
\end{align}
We can actually show that \eqref{eq:LL_lin} is equivalent to \eqref{eq:DGL2} by replacing $S_k^y\rightarrow-\mathrm{i}S_k^y$. Therefore the result of linearized LL-eq. \eqref{eq:LL_lin} corresponds to the quantum mechanical solution when taking the expectation value in a coherent state
\begin{align}
S_k^x(t) &= \langle\alpha|(a_k(t)+a_k(t)^\dagger)|\alpha\rangle, \\
 |\alpha\rangle &= \mathrm{e}^{\alpha a^\dagger-\alpha^*a}|0\rangle
\end{align}
where $\alpha$ is  determined by the initial conditions and $a(t)$ is the time evolution of the bosonic operator $a$ in the Heisenberg picture. This shows that the quantum mechanical and the semi-classical approach are equivalent in the quadratic approximation. The semi-classical ansatz now has the advantage to let us include damping by switching to the LLG eq. \eqref{eq:LLG} which introduces damping phenomenologically in form of an additional term and the damping factor $\alpha$. 
In the linear approximation for small deviations we get for the equations of motion
\begin{align}
\frac{\text{d}}{\text{d}t}\left(\begin{array}{c}
S_k^x \\ S_k^y
\end{array}\right)=
\left(\begin{array}{cc}
\frac{-\alpha(A_k+B_k)}{1+\alpha^2} & \frac{A_k-B_k}{1+\alpha^2}\\
\frac{-A_k-B_k}{1+\alpha^2} & \frac{-\alpha(A_k-B_k)}{1+\alpha^2}
\end{array}\right)
\left(\begin{array}{c}
S_k^x \\ S_k^y
\end{array}\right)
\end{align} 
It should be mentioned that our approximation is valid as long as we are dealing with small numbers of magnons. This enables us to examine the system outside the instability regions and to identify the position of instability regions. In case of resonance our description quickly breaks down due to an exponential rise in magnons, which means that higher order terms like magnon magnon interactions will play a larger role. These are also responsible that the full system will approach a steady state with constant number of magnons \cite{Rezende}.\\
   
\end{document}